\def\tipo{2}      
\ifnum \tipo = 2
\documentstyle[prl,times,twocolumn,aps,epsfig]{revtex} \def\figsize{8cm}
\def\figsiz1{7cm} \def\frontmatter{\twocolumn[ \hsize\textwidth\columnwidth
\hsize\csname@twocolumnfalse\endcsname}  \else
\documentstyle[preprint,times,aps,prl,epsfig]{revtex} \def\figsize{14cm}
\def\figsiz1{12cm} \def\frontmatter{}  \fi

\begin{document}
\draft \frontmatter
\title{Multifractal Detrended Fluctuation Analysis of Nonstationary
Time Series}
\author{Jan W.~Kantelhardt$^{1,2}$, Stephan A.~Zschiegner$^{1}$,
Eva Koscielny-Bunde$^{3,1}$, Armin Bunde$^{1}$, \\
Shlomo Havlin$^{4,1}$, and H. Eugene Stanley$^{2}$}
\address{$^{1}$ Institut f\"ur Theoretische Physik III,
Justus-Liebig-Universit\"at, D-35392 Giessen, Germany}
\address{$^{2}$ Center for Polymer Studies and Department of Physics,
Boston University, Boston MA 02215, USA}
\address{$^{3}$ Potsdam Institute for Climate Impact Research, P.~O.
Box 60 12 03, D-14412 Potsdam, Germany}
\address{$^{4}$ Department of Physics and Gonda-Goldschmied Medical
Diagnostics Research Center, Bar-Ilan University, Ramat-Gan 52900,
Israel}
\maketitle
\begin{abstract}
We develop a method for the multifractal characterization of
nonstationary time series, which is based on a generalization of the
detrended fluctuation analysis (DFA).  We relate our multifractal DFA
method to the standard partition function-based multifractal
formalism, and prove that both approaches are equivalent for
stationary signals with compact support.  By analyzing several
examples we show that the new method can reliably determine the
multifractal scaling behavior of time series.  By comparing the
multifractal DFA results for original series to those for shuffled
series we can distinguish multifractality due to long-range
correlations from multifractality due to a broad probability density
function.  We also compare our results with the wavelet transform
modulus maxima (WTMM) method, and show that the results are equivalent.
\end{abstract}
\bigskip
\ifnum \tipo = 2 ] \fi

\section{Introduction}

In recent years the detrended fluctuation analysis (DFA) method
\cite{Peng94} has become a widely-used technique for the determination
of (mono-) fractal scaling properties and the detection of long-range
correlations in noisy, nonstationary time series
\cite{taqqu,physa,kunhu,kunhu1}.  It has successfully been applied to
diverse fields such as DNA sequences \cite{Peng94,dns}, heart rate
dynamics \cite{herz,PRL00}, neuron spiking \cite{neuron}, human gait
\cite{gait}, long-time weather records \cite{wetter}, cloud structure
\cite{cloud}, geology \cite{malamudjstatlaninfer1999}, ethnology
\cite{Alados2000}, economics time series \cite{economics}, and solid
state physics \cite{fest}.  One reason to employ the DFA method is to
avoid spurious detection of correlations that are artifacts of
nonstationarities in the time series.

Many records do not exhibit a simple monofractal scaling behavior,
which can be accounted for by a single scaling exponent. In some
cases, there exist crossover (time-) scales $s_\times$ separating
regimes with different scaling exponents \cite{physa,kunhu}, e.~g.
long-range correlations on small scales $s \ll s_\times$ and another
type of correlations or uncorrelated behavior on larger scales $s \gg
s_\times$.  In other cases, the scaling behavior is more complicated,
and different scaling exponents are required for different parts of
the series \cite{kunhu1}.  This occurs, e.~g., when the scaling
behavior in the first half of the series differs from the scaling
behavior in the second half.  In even more complicated cases, such
different scaling behavior can be observed for many interwoven fractal
subsets of the time series.  In this case a multitude of scaling
exponents is required for a full description of the scaling behavior,
and a multifractal analysis must be applied.

In general, two different types of multifractality in time series can
be distinguished: (i) Multifractality due to a broad probability
density function for the values of the time series.  In this case the
multifractality cannot be removed by shuffling the series. (ii)
Multifractality due to different long-range (time-) correlations of
the small and large fluctuations. In this case the probability density
function of the values can be a regular distribution with finite
moments, e.~g. a Gaussian distribution.  The corresponding shuffled
series will exhibit non-multifractal scaling, since all long-range
correlations are destroyed by the shuffling procedure. If both kinds
of multifractality are present, the shuffled series will show weaker
multifractality than the original series.

The simplest type of multifractal analysis is based upon the standard
partition function multifractal formalism, which has been developed
for the multifractal characterization of normalized, stationary
measures \cite{feder88,barabasi,peitgen,bacry01}.  Unfortunately, this
standard formalism does not give correct results for nonstationary
time series that are affected by trends or that cannot be normalized. 
Thus, in the early 1990s an improved multifractal formalism has been
developed, the wavelet transform modulus maxima (WTMM) method
\cite{wtmm}, which is based on wavelet analysis and involves tracing
the maxima lines in the continuous wavelet transform over all scales. 
Here, we propose an alternative approach based on a generalization of
the DFA method.  This multifractal DFA (MF-DFA) does not require the
modulus maxima procedure, and hence does not involve more effort in
programming than the conventional DFA.

The paper is organized as follows:  In Section II we describe the
MF-DFA method in detail and show that the scaling exponents determined
via the MF-DFA method are identical to those obtained by the standard
multifractal formalism based on partition functions.  In Section III
we introduce several multifractal models, where the scaling exponents
can be calculated exactly, and compare these analytical results with
the numerical results obtained by MF-DFA.  In Section IV, we show how
the comparison of the MF-DFA results for original series to the MF-DFA
results for shuffled series can be used to determine the type of
multifractality in the series. In Section V, we compare the results of
the MF-DFA with those obtained by the WTMM method for nonstationary
series and discuss the performance of both methods for multifractal
time series analysis.

\section{Multifractal DFA}

\subsection{Description of the method}

The generalized multifractal DFA (MF-DFA) procedure consists of five
steps.  The first three steps are essentially identical to the
conventional DFA procedure (see e.~g.
\cite{Peng94,taqqu,physa,kunhu,kunhu1}).  Let us suppose that $x_k$ is
a series of length $N$, and that this series is of compact support,
i.e. $x_k = 0$ for an insignificant fraction of the values only.

\noindent $\bullet$ {\it Step 1}: Determine the ``profile''
\begin{equation} Y(i) \equiv \sum_{k=1}^i \left[ x_k - \langle x
\rangle \right], \qquad i=1,\ldots,N. \label{profile} \end{equation}
Subtraction of the mean $\langle x \rangle$ is not compulsory, since
it would be eliminated by the later detrending in the third step.

\noindent $\bullet$ {\it Step 2}: Divide the profile $Y(i)$ into $N_s
\equiv {\rm int}(N/s)$ non-overlapping segments of equal length $s$.
Since the length $N$ of the series is often not a multiple of the
considered time scale $s$, a short part at the end of the profile may
remain.  In order not to disregard this part of the series, the same
procedure is repeated starting from the opposite end.  Thereby, $2
N_s$ segments are obtained altogether.

\noindent $\bullet$ {\it Step 3}: Calculate the local trend for each
of the $2 N_s$ segments by a least-square fit of the series.  Then 
determine the variance
\begin{equation} F^2(s,\nu) \equiv {1 \over s} \sum_{i=1}^{s} 
\left\{ Y[(\nu-1) s + i] - y_{\nu}(i) \right\}^2 \label{fsdef} 
\end{equation}
for each segment $\nu$, $\nu = 1, \ldots, N_s$ and
\begin{equation} F^2(s,\nu) \equiv {1 \over s} \sum_{i=1}^{s} 
\left\{ Y[N - (\nu-N_s) s + i] - y_{\nu}(i) \right\}^2 \label{fsdef2}
\end{equation}
for $\nu = N_s+1, \ldots, 2 N_s$.  Here, $y_{\nu}(i)$
is the fitting polynomial in segment $\nu$.  Linear, quadratic, cubic,
or higher order polynomials can be used in the fitting procedure
(conventionally called DFA1, DFA2, DFA3, $\ldots$)
\cite{Peng94,PRL00}. Since the detrending of the time series is done
by the subtraction of the polynomial fits from the profile, different
order DFA differ in their capability of eliminating trends in the
series.  In (MF-)DFA$m$ [$m$th order (MF-)DFA] trends of order $m$ in
the profile (or, equivalently, of order $m - 1$ in the original
series) are eliminated.  Thus a comparison of the results for
different orders of DFA allows one to estimate the type of the
polynomial trend in the time series  \cite{physa,kunhu}.  

\noindent $\bullet$ {\it Step 4}: Average over all segments to obtain
the $q$th order fluctuation function
\begin{equation} F_q(s) \equiv \left\{ {1 \over 2 N_s}
\sum_{\nu=1}^{2 N_s} \left[ F^2(s,\nu) \right]^{q/2} \right\}^{1/q}, 
\label{fdef}\end{equation}
where, in general, the index variable $q$ can take any real value
except zero \cite{note2}.  For $q=2$, the standard DFA procedure is
retrieved.  We are interested in how the generalized $q$ dependent
fluctuation functions $F_q(s)$ depend on the time scale $s$ for
different values of $q$.  Hence, we must repeat steps 2 to 4 for
several time scales $s$.  It is apparent that $F_q(s)$ will increase
with increasing $s$.  Of course, $F_q(s)$ depends on the DFA order $m$. 
By construction, $F_q(s)$ is only defined for $s \ge m+2$.

\noindent $\bullet$ {\it Step 5}: Determine the scaling behavior of
the fluctuation functions by analyzing log-log plots $F_q(s)$ versus
$s$ for each value of $q$.  Several examples of this procedure will be
shown in Section III. If the series $x_i$ are long-range power-law
correlated, $F_q(s)$ increases, for large values of $s$, as a
power-law,
\begin{equation} F_q(s) \sim s^{h(q)} \label{Hq}. \end{equation}
In general, the exponent $h(q)$ may depend on $q$.   For stationary
time series, $h(2)$ is identical to the well-known Hurst exponent $H$
(see e.~g. \cite{feder88}).  Thus, we will call the function $h(q)$
generalized Hurst exponent.  

For monofractal time series with compact support, $h(q)$ is
independent of $q$, since the scaling behavior of the variances
$F^2(s,\nu)$ is identical for all segments $\nu$, and the averaging
procedure in Eq.~(\ref{fdef}) will give just this identical scaling
behavior for all values of $q$.  Only if small and large fluctuations
scale differently, there will be a significant dependence of $h(q)$ on
$q$:  If we consider positive values of $q$, the segments $\nu$ with
large variance $F^2_s(\nu)$ (i.~e. large deviations from the
corresponding fit) will dominate the average $F_q(s)$.  Thus, for
positive values of $q$, $h(q)$ describes the scaling behavior of the
segments with large fluctuations.  Usually the large fluctuations are
characterized by a smaller scaling exponent $h(q)$ for multifractal
series \cite{note3}.  On the contrary, for negative values of $q$, the
segments $\nu$ with small variance $F^2_s(\nu)$ will dominate the
average $F_q(s)$.  Hence, for negative values of $q$, $h(q)$ describes
the scaling behavior of the segments with small fluctuations, which
are usually characterized by a larger scaling exponent.

However, the MF-DFA method can only determine {\it positive}
generalized Hurst exponents $h(q)$, and it already becomes inaccurate
for strongly anti-correlated signals when $h(q)$ is close to zero. In
such cases, a modified (MF-)DFA technique has to be used.  The most
simple way to analyze such data is to integrate the time series before
the MF-DFA procedure.  Hence, we replace the {\it single} summation in
Eq.~(\ref{profile}), which is describing the profile from the original
data $x_k$, by a {\it double} summation,
\begin{equation} \tilde{Y}(i) \equiv \sum_{k=1}^i \left[ Y(k) -
\langle Y \rangle \right]. \label{double} \end{equation}
Following the MF-DFA procedure as described above, we obtain
generalized fluctuation functions $\tilde{F}_q(s)$ described by a
scaling law as in Eq.~(\ref{Hq}), but with larger exponents
$\tilde{h}(q) = h(q) + 1$,
\begin{equation} \tilde{F}_q(s) \sim s^{\tilde{h}(q)} = s^{h(q)+1}.
\label{Htilde} \end{equation}
Thus, the scaling behavior can be accurately determined even for
$h(q)$ which are smaller than zero (but larger than $-1$) for some
values of $q$.  We note that $\tilde{F}_q(s)/s$ corresponds to
$F_q(s)$ in Eq.~(\ref{Hq}).  If we do not subtract the average values
in each step of the summation in Eq.~(\ref{double}), this summation
leads to quadratic trends in the profile $\tilde{Y}(i)$.  In this case
we must employ at least the second order MF-DFA to eliminate these
artificial trends.

\subsection{Relation to standard multifractal analysis}

For stationary, normalized records with compact support the
multifractal scaling exponents $h(q)$ defined in Eq.~(\ref{Hq}) are
directly related, as shown below, to the scaling exponents $\tau(q)$
defined by the standard partition function-based multifractal
formalism.

Suppose that the series $x_k$ of length $N$ is a stationary,
normalized sequence.  Then the detrending procedure in step 3 of the 
MF-DFA method is not required, since no trend has to be eliminated.
Thus, the DFA can be replaced by the standard Fluctuation Analysis
(FA), which is identical to the DFA except for a simplified definition
of  the variance for each segment $\nu$, $\nu = 1, \ldots, N_s$, in
step 3 [see Eq.~(\ref{fsdef})]:
\begin{equation} F_{\rm FA}^2(s,\nu) \equiv [Y(\nu s) - Y((\nu-1) s)]^2.
\label{FAfsdef} \end{equation}
Inserting this simplified definition into Eq.~(\ref{fdef}) and using
Eq.~(\ref{Hq}), we obtain
\begin{equation} \left\{ {1 \over 2 N_s} \sum_{\nu=1}^{2 N_s} 
\vert Y(\nu s) - Y((\nu-1) s) \vert^q \right\}^{1/q} \sim s^{h(q)}. 
\label{FAfHq} \end{equation}
For simplicity we can assume that the length $N$ of the series is an
integer multiple of the scale $s$, obtaining $N_s = N/s$ and
therefore
\begin{equation} \sum_{\nu=1}^{N/s} \vert Y(\nu s) - Y((\nu-1) s)
\vert^q \sim s^{q h(q) - 1}. \label{MFA} \end{equation}
This already corresponds to the multifractal formalism used e.~g. in
\cite{barabasi,bacry01}.  In fact, a hierarchy of exponents $H_q$
similar to our $h(q)$ has been introduced based on Eq.~(\ref{MFA})
in \cite{barabasi}. 

In order to relate also to the standard textbook box counting
formalism \cite{feder88,peitgen}, we employ the definition of the
profile in Eq.~(\ref{profile}).  It is evident that the term
$Y(\nu s) - Y((\nu-1) s)$ in Eq.~(\ref{MFA}) is identical to the
sum of the numbers $x_k$ within each segment $\nu$ of size $s$. 
This sum is known as the box probability $p_s(\nu)$ in the standard
multifractal formalism for normalized series $x_k$,
\begin{equation} p_s(\nu) \equiv \sum_{k=(\nu-1) s +1}^{\nu s} x_k =
Y(\nu s) - Y((\nu-1) s).  \label{boxprob} \end{equation}
The scaling exponent $\tau(q)$ is usually defined via the partition
function $Z_q(s)$,
\begin{equation} Z_q(s) \equiv \sum_{\nu=1}^{N/s} \vert p_s(\nu) 
\vert^q \sim s^{\tau(q)}, \label{Zq} \end{equation}
where $q$ is a real parameter as in the MF-DFA above.  Sometimes 
$\tau(q)$ is defined with opposite sign (see e.~g. \cite{feder88}).

Using Eq.~(\ref{boxprob}) we see that Eq.~(\ref{Zq}) is identical to
Eq.~(\ref{MFA}), and obtain analytically the relation between the two
sets of multifractal scaling exponents,
\begin{equation} \tau(q) = q h(q) - 1. \label{tauH} \end{equation}
Thus, we have shown that $h(q)$ defined in Eq.~(\ref{Hq}) for the
MF-DFA is directly related to the classical multifractal scaling 
exponents $\tau(q)$.  Note that $h(q)$ is different from the 
generalized multifractal dimensions
\begin{equation} D(q) \equiv {\tau(q) \over q-1} =
{q h(q)-1 \over q-1}, \label{Dq} \end{equation}
that are used instead of $\tau(q)$ in some papers.  While $h(q)$ is
independent of $q$ for a monofractal time series with compact support,
$D(q)$ depends on $q$ in that case.  Our assumption of compact support
of the series $x_k$ can be directly observed in Eq.~(\ref{Dq}), since
the fractal dimension of the support is $D(0) \equiv -\tau(0) = 1$. 

Another way to characterize a multifractal series is the singularity
spectrum $f(\alpha)$, that is related to $\tau(q)$ via a Legendre
transform \cite{feder88,peitgen},
\begin{equation} \alpha = \tau'(q) \quad {\rm and} \quad
f(\alpha) = q \alpha - \tau(q). \label{Legendre} \end{equation}
Here, $\alpha$ is the singularity strength or H\"older exponent, while
$f(\alpha)$ denotes the dimension of the subset of the series that is
characterized by $\alpha$. Using Eq.~(\ref{tauH}), we can directly
relate $\alpha$ and  $f(\alpha)$ to $h(q)$, 
\begin{equation} \alpha = h(q) + q h'(q) \quad {\rm and} \quad
f(\alpha) = q [\alpha - h(q)] + 1.\label{Legendre2} \end{equation}

\section{Four Illustrative examples}

\subsection{Example 1: monofractal uncorrelated and long-range
correlated series}

As a first example we apply the MF-DFA method to monofractal series
with compact support, for which the generalized Hurst exponent $h(q)$
is expected to be independent of $q$,
\begin{equation} h(q) = H \quad {\rm and} \quad \tau(q) = q H -
1. \label{mono}\end{equation}
Such series have been discussed in the context of conventional DFA in
several studies before, see e.g. \cite{physa,kunhu,kunhu1}. Long-range
correlated random numbers are usually generated by the Fourier
transform method, see e.~g. \cite{feder88,makse96}.  Using this method
we can generate long-range anti-correlated ($0 < H < 0.5$),
uncorrelated ($H = 0.5$), or (positively) long-range correlated ($0.5
< H < 1$) series.  The latter are characterized by a power-law decay
of the autocorrelation function $C(s) \equiv \langle x_k \, x_{k+s}
\rangle \sim s^{-\gamma}$ for large scales $s$ with $\gamma = 2 - 2 H$
if the series is stationary.  Alternatively, all stationary long-range
correlated series can be characterized by the power-law decay of their
power spectra, $S(f) \sim f^{-\beta}$ with frequency $f$ and $\beta =
2 H - 1$.  Note that $H$ corresponds to the Hurst exponent of the
integrated time series here.

\begin{figure}\centering
\epsfxsize\figsize\epsfbox{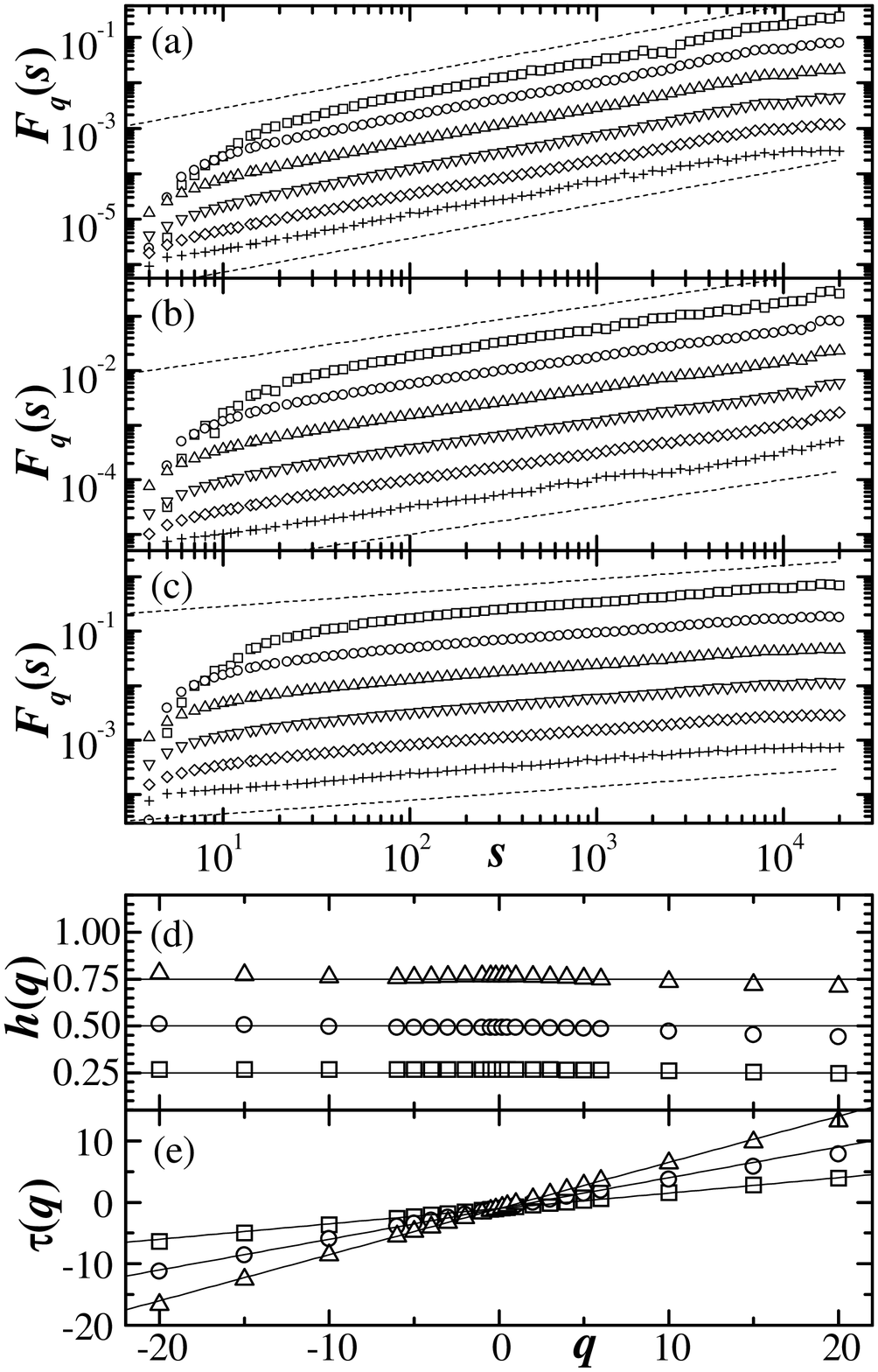}\vspace{1cm}

\parbox{\figsize}{\caption[]{\small
The MF-DFA fluctuation functions $F_q(s)$ are shown versus the scale
$s$ in log-log plots for (a) long-range correlated monofractal series
with $H=0.75$, (b) uncorrelated random series with $H=0.5$ (white
noise), and (c) long-range anti-correlated series with $H=0.25$. The
different symbols correspond to the different values of the exponent
$q$ in the generalized averaging procedure, $q=-10$ ($\Box$), $-2$
({\Large$\circ$}), $-0.2$ ($\bigtriangleup$), $+0.2$
($\bigtriangledown$), $+2$ ($\Diamond$), and $q=+10$ ($+$). MF-DFA2 has
been employed, and the curves have been shifted by multiple factors of
$4$ for clarity.  The straight dashed lines have the corresponding
slopes $H$ and are shown for comparison.  Part (d) shows the $q$
dependence of the asymptotic scaling exponent $h(q)$ determined by
fits in the regime $200 < s < 5000$ for $H=0.25$ ($\Box$), 0.5
({\Large$\circ$}), and $0.75$ ($\bigtriangleup$). The very weak
dependence on $q$ is consistent with monofractal scaling.  In part
(e) the $q$ dependence of $\tau(q)$, $\tau(q)=q h(q)-1$, is shown.
\label{mono1}}}\end{figure}

Figure \ref{mono1} shows the generalized fluctuation functions
$F_q(s)$ for all three types of monofractal series ($H=0.75, 0.5,
0.25$) and several $q$ values.  On large scales $s$, we observe the
expected power-law scaling behavior according to Eq.~(\ref{Hq}), which
corresponds to straight lines in the log-log plot.  In
Fig.~\ref{mono1}(d), the scaling exponents $h(q)$ determined from the
slopes of these straight lines are shown versus $q$.  Although a
slight $q$ dependence is observable, the values of $h(q)$ are always
very close to the $H$ of the generated series that has been
analyzed.  The degree of the $q$ dependence observed for this
monofractal series allows to estimate the usual fluctuation of $h(q)$
to be expected for monofractal series in general.  

Next, we analyze multifractal series for which $\tau(q)$ can be
calculated exactly, and compare the numerical results with the
expected scaling behavior.

\subsection{Example 2: binomial multifractal series}

In the binomial multifractal model \cite{feder88,barabasi,peitgen}, a
series of $N = 2^{n_{\rm max}}$ numbers $k$ with $k=1, \ldots, N$ is
defined by
\begin{equation} x_k = a^{n(k-1)} (1-a)^{n_{\rm max}-n(k-1)},
\label{bindef} \end{equation}
where $0.5 < a < 1$ is a parameter and $n(k)$ is the number of digits
equal to 1 in the binary representation of the index $k$, e.~g. $n(13)
= 3$, since 13 corresponds to binary 1101.

The scaling exponents $\tau(q)$ can be calculated straightforwardly. 
According to Eqs.~(\ref{boxprob}) and (\ref{bindef}) the box
probability $p_{2s}(\nu)$ in the $\nu$th segment of size $2s$ is given
by
\begin{eqnarray*} p_{2s}(\nu) = p_{s}(2 \nu - 1) + p_{s}(2 \nu) \\
= [(1-a)/a + 1] p_{s}(2 \nu) = p_{s}(2\nu) /a. \end{eqnarray*}
Thus, according to Eqs.~(\ref{Zq}) and (\ref{bindef}),
\begin{eqnarray*} Z_q(s) = \sum_{\nu=1}^{N/s} [p_s(\nu)]^q
= \sum_{\nu=1}^{N/2s} [p_{s}(2 \nu - 1)]^q + [p_{s}(2 \nu)]^q \\
= \left[ {(1-a)^q \over a^q} + 1 \right] \sum_{\nu=1}^{N/2s}
[p_{s}(2 \nu)]^q \\
= [(1-a)^q + a^q] \sum_{\nu=1}^{N/2s} [p_{2s}(\nu)]^q
= [(1-a)^q + a^q] \; Z_q(2s) \end{eqnarray*}
and according to Eqs.~(\ref{Zq}) and (\ref{tauH}),
\begin{eqnarray} \tau(q) = {-\ln[a^q + (1-a)^q] \over \ln(2)},
\label{taubin} \\
h(q) = {1 \over q} - { \ln[a^q + (1-a)^q] \over q\ln(2)}.
\label{Hbin} \end{eqnarray}

Note that $\tau(0)=-1$ as required.  There is a strong non-linear
dependence of $\tau(q)$ upon $q$, indicating multifractality. The same
information is comprised in the $q$ dependence of $h(q)$. The
asymptotic values are $h(q) \to -\ln(a)/\ln(2)$ for $q \to +\infty$
and $h(q) \to -\ln(1-a)/\ln(2)$ for $q \to -\infty$.  They correspond
to the scaling behavior of the largest and weakest fluctuations,
respectively.  Note that $h(q)$ becomes independent of $q$ in the
asymptotic limit, while $\tau(q)$ approaches linear $q$ dependences.

\begin{figure}\centering
\epsfxsize\figsize\epsfbox{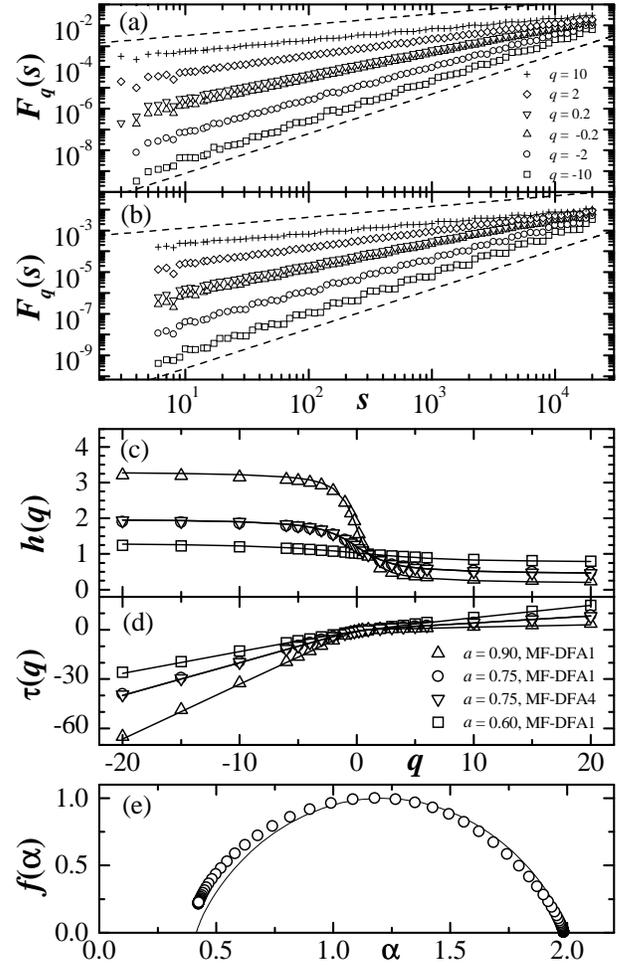}\vspace{1cm}

\parbox{\figsize}{\caption[]{\small
The MF-DFA fluctuation functions $F_q(s)$ are shown versus the scale
$s$ in log-log plots for the binomial multifractal model with $a=0.75$
(a) for MF-DFA1 and (b) for MF-DFA4.  The symbols are the same as for
Fig.~1.  The straight dashed lines have the corresponding theoretical
slopes $h(-10)=1.90$ and $h(+10)=0.515$ and are shown for comparison. 
In part (c) the $q$ dependence of the generalized Hurst exponent
$h(q)$ determined by fits in the regime $50 < s < 500$ is shown for
MF-DFA1 and  $a=0.9$ ($\bigtriangleup$), $a=0.75$ ({\Large$\circ$}),
and $a=0.6$ ($\Box$), as well as for MF-DFA4 and $a=0.75$
($\bigtriangledown$).  Parts (d) and (e) show the corresponding
exponents $\tau(q)$ and the corresponding singularity spectrum
$f(\alpha)$ for $a=0.75$ determined by the modified Legendre transform
(\protect\ref{Legendre2}), respectively.  The lines are the
theoretical values obtained from Eq.~(\protect\ref{Hbin}).
\label{bin1}}}\end{figure}

Figure \ref{bin1} shows the MF-DFA fluctuation functions $F_q(s)$ for
the binomial multifractal model with $a=0.75$.  The results for
MF-DFA1 and MF-DFA4 are compared in parts (a) and (b).  Fig.
\ref{bin1}(c) shows the corresponding slopes $h(q)$ for three values
of $a$ together with the exact results obtained from
Eq.~(\ref{Hbin}).  The numerical results are in good agreement with
Eq.~(\ref{Hbin}), showing that the MF-DFA correctly detects the
multifractal scaling exponents. Figures~\ref{bin1}(d) and (e) show the
corresponding exponents $\tau(q) = q h(q) - 1$ [see Eq.~(\ref{tauH})]
and the corresponding $f(\alpha)$ spectrum calculated from $h(q)$
using the modified Legendre transform (\ref{Legendre2}).  Both are
also in good agreement with Eq.~(\ref{taubin}). We have also checked
that the results for the binomial multifractal model remain unchanged
if the double summation technique [see Eq.~(\ref{double})] is
applied.  We obtain slopes $\tilde{h}(q) = h(q) + 1$ as expected in
Eq.~(\ref{Htilde}).  Note that there is no need to use this
modification, except if $h(q)$ is close to zero or has negative
values.

\subsection{Example 3: dyadic random cascade model with log-Poisson
distribution}

For another independent test of the MF-DFA, we employ an algorithm
based on random cascades on wavelet dyadic trees proposed in
\cite{arneodo98} (see also \cite{PREprep}).  This algorithm builds a
random multifractal series by specifying its discrete wavelet
coefficients $c_{n,m}$, defined recursively,
\begin{eqnarray*} c_{1,1} = 1, \quad c_{n,2m-1} = W c_{n-1,m},
\quad c_{n,2m} = W c_{n-1,m}, \end{eqnarray*}
where $n=2,\ldots,n_{\rm max}$ (with $N=2^{n_{\rm max}}$) and
$m=1,\ldots,2^{n-2}$.  The values of $W$ are taken from a log-Poisson
distribution, $\vert W \vert = \exp( P \ln \delta + \gamma )$, where
$P$ is Poisson distributed with $\langle P \rangle = \lambda$. There
are three independent parameters, $\lambda$, $\delta$, and $\gamma$.
Inverse wavelet transform is applied to create the multifractal random
series $x_k$ once the wavelet coefficients $c_{n,m}$ are known,
\begin{equation} x_k = \sum_{n=1}^{n_{\rm max}}
\sum_{m=1}^{2^{n-1}} c_{n,m} \psi_{n,m}(k), \end{equation}
where $\psi_{n,m}(k)$ is a set of wavelets forming an orthonormal
wavelet basis.  Here, we employ the Haar wavelets,  $\psi_{n,m}(k)
\equiv 2^{(n-n_{\rm max}-1)/2} \psi[ 2^{n-n_{\rm max}-1} k - m]$ with
$\psi(x) \equiv 1$ for $0 < x \le 0.5$, $\psi(x) \equiv -1$ for $0.5 <
x \le 1$ and $\psi(x) \equiv 0$ otherwise. For this model the
multifractal scaling exponents are given by \cite{arneodo98}
\begin{eqnarray} \tau(q) = {\lambda (1 - \delta^q) - \gamma q
\over \ln 2} -1, \\ h(q) = [\lambda (1 - \delta^q) - \gamma q]
/ (q \ln 2). \label{Hdya} \end{eqnarray}

\begin{figure}\centering
\epsfxsize\figsize\epsfbox{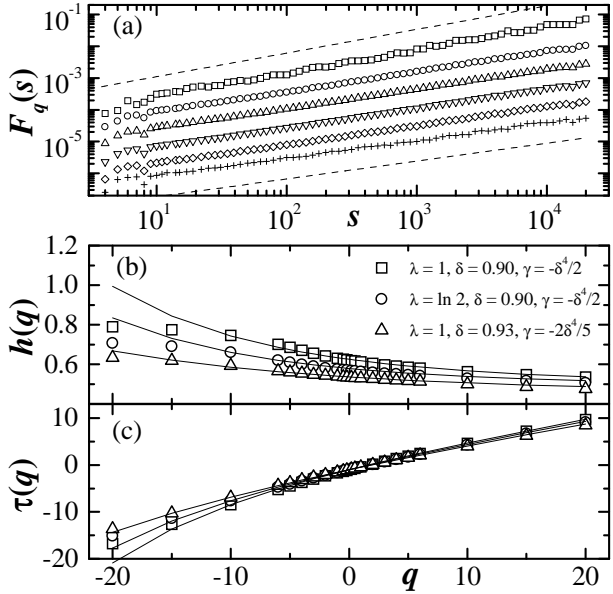}\vspace{1cm}

\parbox{\figsize}{\caption[]{\small
(a) The MF-DFA2 fluctuation functions $F_q(s)$ are shown versus the
scale  $s$ in a log-log plot for the dyadic random cascade model with
log-Poisson distribution with parameters $\lambda = 1$, $\delta =
0.9$, and $\gamma = -\delta^4/2$.  The symbols are the same as for
Fig.~1. The dashed straight lines have the theoretical slopes $h(-10)
= 0.743$ and $h(+10) = 0.567$ and are shown for comparison. (b) The $q$
dependence of the generalized Hurst exponent $h(q)$ determined by fits
is shown for MF-DFA2 and different parameters (see legend).  The lines
are the theoretical values obtained from Eq.~(\protect\ref{Hdya}).  In
part (c) $\tau(q)=q h(q)-1$ is shown. \label{dya1}}}\end{figure}

Figure \ref{dya1} shows the MF-DFA fluctuation functions $F_q(s)$ for
the dyadic random cascade model.  The numerically determined slopes
$h(q)$ for three sets of parameters are compared with the exact
results obtained from Eq.~(\ref{Hdya}) and the good agreement shows
that the MF-DFA correctly detects the multifractal scaling exponents. 
Large deviations occur only for very small moments ($q<-10$),
indicating that the range of $q$ values should not exceed $-10$.

\subsection{Example 4: uncorrelated multifractal series with
power-law distribution function}

The examples discussed in the previous three subsections were based on
series involving long-range correlations.  In the present example we
want to apply the MF-DFA method to an uncorrelated series, that
nevertheless exhibits multifractal scaling behavior due to the broad
distribution of its values.  We denote by $P(x)$ the probability
density function of the values $x_k$ in the series.  The distribution
$P(x)$ does not affect the multifractality of a series on large scales
$s$, if all moments
\begin{equation} \langle \vert x \vert^q \rangle \equiv
\int_{-\infty}^\infty \vert x \vert^q P(x) \, dx \end{equation}
are finite.  Here we choose a (normalized) power-law probability
distribution function,
\begin{equation} P(x) = \alpha x^{-(\alpha+1)} \quad {\rm for} 
\; 1 \le x < \infty \quad {\rm with} \; \alpha > 0
\label{Pvonx} \end{equation}
and $P(x)=0$ for $x<1$, where already the second moment diverges
if $\alpha \le 2$.  In this
case, the series exhibits multifractal scaling behavior on all
scales.  Note, that Eq.~(\ref{Pvonx}) becomes identical to a Levy
distribution of class $\alpha$ for large values of $x$.  The parameter
$\alpha$ is not related to the H\"older exponent $\alpha$ in
Eq.~(\ref{Legendre}).  The scaling properties of random walks with
Levy distributed steps (Levy flights and Levy walks) have been
analyzed in \cite{shle87,havlin99,newprep}.  The multifractal nature
of Levy processes has been investigated in \cite{jaffard,nakao}.

In order to derive the multifractal spectrum, let us consider $s$
uncorrelated random numbers $r_k$, $k=1,\ldots,s$, distributed
homogeneously in the interval $[0,1]$.  Obviously, the typical value of
the minimum of the numbers, $r_{\rm min}(s) \equiv {\rm min}_{k=1}^s
r_k$, will be $r_{\rm min}(s) = 1/s$.  It can be easily shown that the
numbers $r_k$ are transformed into numbers $x_k$ distributed
according to the power-law probability distribution function
(\ref{Pvonx}) by $r_k \to x_k = r_k^{-1/\alpha}$.  Thus, the typical
value of the maximum of the $x_k$ will be $x_{\rm max}(s) \equiv {\rm
max}_{k=1}^s x_k = [r_{\rm min}(s)]^{-1/\alpha} = s^{1/\alpha}$.  

If $\alpha \le 2$, the fluctuations of the profile $Y(i)$
[Eq.~(\ref{profile})] and the corresponding DFA variance $F^2(s,\nu)$
[Eq.~(\ref{fsdef})] will be dominated by the square of the largest
value $x^2_{\rm max}(s) = s^{2/\alpha}$ in the segment of $s$ numbers,
since the second moment of the distribution (\ref{Pvonx}) diverges. 
Now the whole series consists of $N_s \equiv {\rm int}(N/s)$ segments
of length $s$ and not just of one segment.  For some segments $\nu$,
$[F^2(s,\nu)]^{1/2}$ is larger than its typical value $x_{\rm max}(s)
= s^{1/\alpha}$, since the maximum within the whole series of length
$N$ is $x_{\rm max}(N) = N^{1/\alpha}$.   In order to calculate
$F_q(s)$ [Eq.~(\ref{fdef})], we need to take into account the whole
distribution $P_s(y)$ of the values $y \equiv [F^2(s,\nu)]^{1/2}$. 
Since each of the maxima in the $N_s$ segments corresponds to an
actual number $x_k$ and these $x_k$ are random numbers from the
power-law distribution (\ref{Pvonx}), it becomes obvious, that the
distribution of the maxima will have the same form, i.~e. $P_s(y) \sim
P(x=y)$ for large $y$.  Small values of $y$ are excluded because of
the maximum procedure, but the large $x_k$ values are very likely to
be identical to the maxima of the corresponding segments.  Since the
smallest maxima for segments of length $s$ are of the order of $x_{\rm
max}(s) = s^{1/\alpha}$, the lower cutoff for $P_s(y)$ must be
proportional to $s^{1/\alpha}$.  From the normalization condition
$\int_{A s^{1/\alpha}}^\infty P_s(y) \; dy = 1$ (with an unimportant
prefactor $A<1$) we get
\begin{equation} P_s(y) = A^\alpha \alpha s y^{-(\alpha+1)}. 
\end{equation}

Now $F_q(s)$ [Eq.~(\ref{fdef})] can be calculated by integration from
the minimum value $A s^{1/\alpha}$ of $y \equiv [F^2(s,\nu)]^{1/2}$
to the maximum value $N^{1/\alpha}$.  For $s \ll N$ we obtain
\begin{eqnarray*} F_q(s) \sim \left[
\int_{As^{1/\alpha}}^{N^{1/\alpha}} y^q P_s(y) \; dy \right]^{1/q}
\\ \sim \left \vert A^\alpha s N^{q/\alpha - 1} - A^q s^{q/\alpha}
\right \vert^{1/q} \sim \cases{ s^{1/q} \quad (q > \alpha)
\cr s^{1/\alpha} \quad (q < \alpha) }. \end{eqnarray*}
Comparing with Eq.~(\ref{Hq}), we finally get
\begin{equation} h(q) \sim \cases{ 1/q \quad (q > \alpha) \cr
1/\alpha \quad (q \le \alpha) }. \label{Hbroa} \end{equation}
Note that $\tau(q)$ follows a linear $q$ dependence, $\tau(q) =
q/\alpha -1$ for $q < \alpha$, while it is equal to zero for  $q >
\alpha$ according to Eq.~(\ref{tauH}).  Hence, the series of
uncorrelated power-law distributed values has rather bi-fractal
\cite{nakao}
instead of multifractal properties.  Since $h(2) = 1/2$ holds exactly
for all values of $\alpha$, it is not possible to recognize the
multifractality due to the broad power-law distribution of the values
if only the conventional DFA is applied.  The second moment shows just the
uncorrelated behavior of the values.  In a very recent preprint
\cite{newprep} this behavior has been interpreted as a failure of the
DFA and corresponding non-detrending methods for series with a broad
distribution, and another method to determine the exponent $1/\alpha$
has been proposed.  We believe that a multifractal description with
more than one exponent is required to characterize this kind of
series, and thus any method calculating just one exponent will be
insufficient for a full characterization. 

\begin{figure}\centering
\epsfxsize\figsize\epsfbox{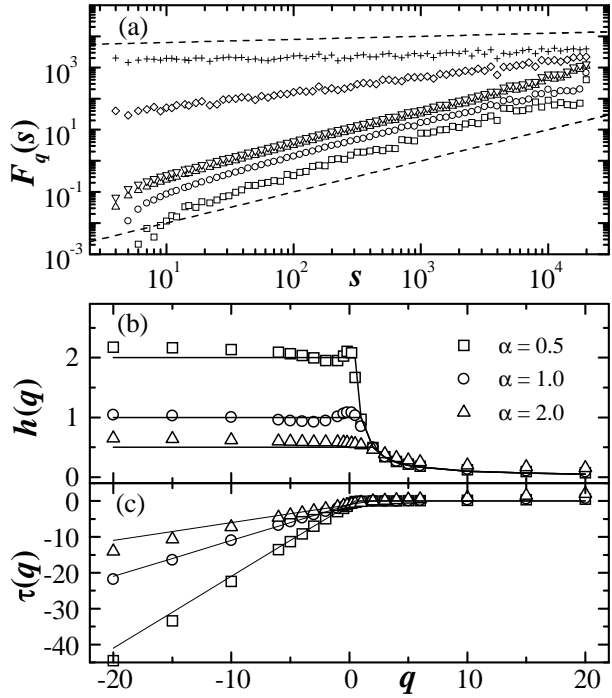}\vspace{1cm}

\parbox{\figsize}{\caption[]{\small
(a) The modified and rescaled MF-DFA3 fluctuation functions
$\tilde{F}_q(s)/s [\cong F_q(s)]$ are shown versus the scale $s$ in a
log-log plot for a series of independent numbers with a power-law
probability density distribution $P(x) \sim x^{-(\alpha+1)}$ with
$\alpha=1$.  The symbols are the same as for Fig.~1.  The straight
dashed lines have the corresponding theoretical slopes $h(-10)=1$
and $h(+10)=0.1$ and are shown for comparison.  (b) The $q$ dependence
of the generalized Hurst exponent $h(q) = \tilde{h}(q)-1$ determined
by fits on large scales $s$ is shown for MF-DFA3 and $\alpha =0.5$
($\Box$), $1.0$ ({\Large$\circ$}), and $2.0$ ($\bigtriangleup$).  The
lines are the theoretical values obtained from
Eq.~(\protect\ref{Hbroa}).  In part (c) the corresponding $\tau(q)$ is
shown.  The broad distribution of the values leads to multifractality
(bi-fractality) in all three cases. \label{broa1}}}\end{figure}

Figure \ref{broa1}(a) shows the MF-DFA3 fluctuation functions for
series of independent random numbers $x_k \in [1,\infty)$ distributed
according to Eq.~(\ref{Pvonx}) with $\alpha=1$.  Since the scaling
exponents $h(q)$ become very close to zero asymptotically for large
positive values of $q$ according to Eq.~(\ref{Hbroa}), we must use the
modified MF-DFA technique involving the double sum as described in the
last paragraph of Subsection II.A.  Hence, for this technical reason,
$\tilde{F}_q(s)/s$ is calculated instead of $F_q(s)$.  The
corresponding slopes $\tilde{h}(q)-1$ are identical to $h(q)$, see
Eq.~(\ref{Htilde}).  In Fig.~\ref{broa1}(b) the slopes $h(q)$ for
series with $\alpha=0.5$, 1.0, and 2.0 are compared with the
theoretical result Eq.~(\ref{Hbroa}), and nice agreement is observed.

\section{Comparison of the multifractality for original and
shuffled series}

\subsection{Distinguishing the two types of multifractality}

As already mentioned in the introduction, two different types of
multifractality in time series can be distinguished.  Both of them
require a multitude of scaling exponents for small and large
fluctuations.  (i) Multifractality of a time series can be due to a
broad probability density function for the values of the time series,
and (ii) multifractality can also be due to different long-range
correlations for small and large fluctuations.  The example
discussed in Subsection III.D, the uncorrelated multifractal series
with a power-law probability density function, is of type (i), while
the examples discussed in Subsections III.A -- III.C are of type (ii),
where the probability density function of the values is a regular
distribution with finite moments \cite{note4}. 

Now, we would like to distinguish between these two types of
multifractality.  The most easy way to do so is by analyzing also the 
corresponding randomly shuffled series.  In the shuffling procedure
the values are put into random order, and thus all correlations are
destroyed.  Hence the shuffled series from multifractals of type (ii)
will exhibit simple random behavior, $h_{\rm shuf}(q) = 0.5$, i.~e.
non-multifractal scaling like in Fig.~\ref{mono1}(b).  For
multifractals of type (i), on the contrary, the original $h(q)$
dependence is not changed,  $h(q)=h_{\rm shuf}(q)$, since the
multifractality is due to the probability density, which is not
affected by the shuffling procedure.  If both kinds of multifractality
are present in a given series, the shuffled series will show weaker
multifractality than the original one.

\begin{figure}\centering
\epsfxsize\figsize\epsfbox{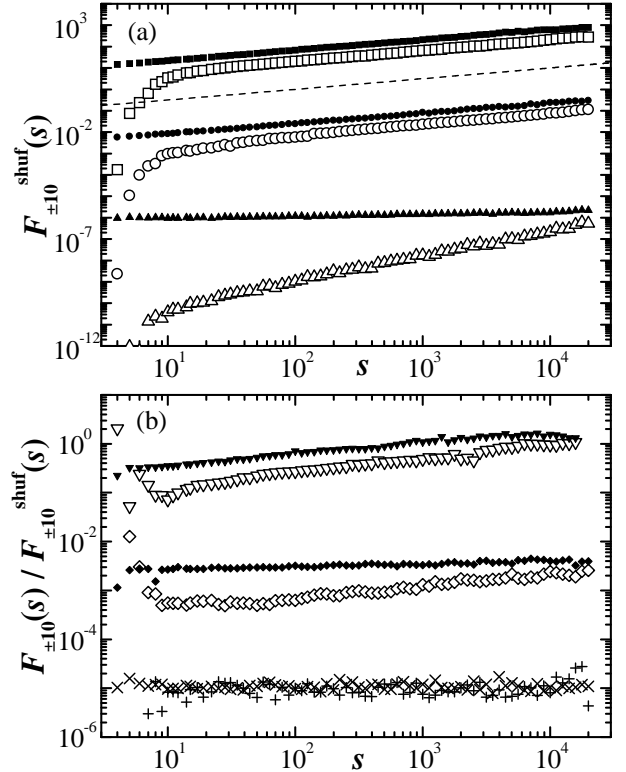}\vspace{1cm}

\parbox{\figsize}{\caption[]{\small
(a) The MF-DFA fluctuation functions $F_{-10}^{\rm shuf}(s)$ (open
symbols) and $F_{10}^{\rm shuf}(s)$ (filled symbols) are shown versus
the scale $s$ in a log-log plot for randomly shuffled series of
long-range correlated series with $H=0.75$ ($\Box$), for the dyadic
random cascade model with log-Poisson distribution with parameters
$\lambda = \ln 1$, $\delta = 0.9$, and $\gamma = -\delta^4/2$
({\Large$\circ$}), and for power-law distributed random numbers $x_k$
with $P(x)\sim x^{-2}$ ($\bigtriangleup$).  The correlations and the
multifractality are destroyed by the shuffling procedure for the first
two series, but for the broadly distributed random numbers the
multifractality remains. The dashed line has the slope $H=0.5$ and is
shown for comparison. (b) The ratios of the MF-DFA2 fluctuation
functions $F_q(s)$ of the original series and the MF-DFA2 fluctuation
functions $F_q^{\rm shuf}(s)$ of the randomly shuffled series are
shown versus $s$ for the same models as in (a), correlated series
($\bigtriangledown$), dyadic random cascade model ($\Diamond$), and
power-law distributed random numbers ($+$ for $q=-10$, $\times$ for
$q=+10$).  The deviations from the slope $h_{\rm cor} = 0$ indicate
long-range correlations. \label{shuf1}}}\end{figure}

\begin{figure}\centering
\epsfxsize\figsize\epsfbox{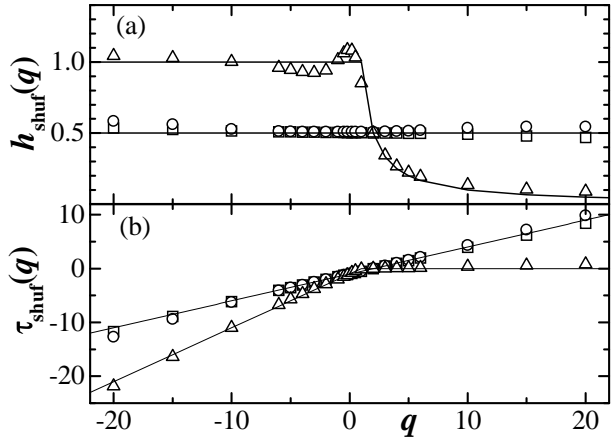}\vspace{1cm}

\parbox{\figsize}{\caption[]{\small
(a) The $q$ dependence of the slopes $h_{\rm shuf}(q)$ of the same
models as in Fig.~\protect{\ref{shuf1}}(a).  The lines indicate the
theoretical values:  $H=0.5$ for shuffled data with narrow
distribution, and $h(q)$ from Eq.~(\protect\ref{Hbroa}) for the series
of numbers with a power-law probability density distribution.  The
symbols are the same as in Fig.~\protect{\ref{shuf1}}.  Part (b) shows
$\tau_{\rm shuf}(q)=q h_{\rm shuf}(q) -1$. \label{shuf2}}}
\end{figure}

The effect of the shuffling procedure is illustrated in
Fig.~\ref{shuf1}(a), where the MF-DFA2 fluctuation functions
$F_{-10}^{\rm shuf}(s)$ and $F_{10}^{\rm shuf}(s)$ are shown for
shuffled series for three of the multifractal examples taken from the
previous section.  Random behavior, $h_{\rm shuf}(q) = 0.5$, is
observed for the series that were long-range correlated or generated
from the dyadic random cascade model before the shuffling procedure
[upper four curves in Fig.~\ref{shuf1}(a)].  In contrast, we observe
the original multifractal scaling for the shuffled multifractal series
with power-law probability density function $P(x)\sim x^{-2}$ [lower
two curves in Fig.~\ref{shuf1}(a)].  The $h_{\rm shuf}(q)$ dependences
are shown in Fig.~\ref{shuf2}, which can be compared with the
corresponding slopes shown in Figs.~\ref{mono1}(d), \ref{dya1}(b), and
\ref{broa1}(b).  Thus, the fluctuation analysis of the shuffled
series, $F_q^{\rm shuf}(s)$, directly indicates the presence of type
(i) multifractality, which is due to a broad probability distribution,
by deviations from $h_{\rm shuf}(q)=0.5$.

Now we want to determine directly the magnitude of the (ii) 
multifractality, which is due to correlations.  For that purpose we
compare the fluctuation function for the original series, $F_q(s)$,
with the result for the corresponding shuffled series, $F_q^{\rm
shuf}(s)$.  Differences between these two fluctuation functions
directly indicate the presence of correlations in the original
series.  These differences can be observed best in a plot of the ratio
$F_q(s) / F_q^{\rm shuf}(s)$ versus $s$ \cite{note5}.  Since the
anomalous scaling due to a broad probability density affects $F_q(s)$
and $F_q^{\rm shuf}(s)$ in the same way, only multifractality due to
correlations will be observed in $F_q(s) / F_q^{\rm shuf}(s)$.  This
is illustrated in Fig.~\ref{shuf1}(b) for the same three multifractal
examples as in Fig.~\ref{shuf1}(a).  In order not to have increased
statistical errors in the results when considering the ratio $F_q(s)
/ F_q^{\rm shuf}(s)$ instead of $F_q(s)$ itself, $F_q^{\rm shuf}(s)$
can be calculated by averaging over a large number of randomly
shuffled series generated from the same original series.

The scaling behavior of the ratio is
\begin{equation} F_q(s) / F_q^{\rm shuf}(s) \sim s^{h(q)-h_{\rm
shuf}(q)} = s^{h_{\rm cor}(q)}. \label{HqCor} \end{equation}
Note that $h(q)=h_{\rm shuf}(q)+h_{\rm cor}(q)$.  If only distribution
multifractality [type (i)] is present, $h(q)=h_{\rm shuf}(q)$ depends
on $q$ and $h_{\rm cor}(q)=0$.  On the other hand, deviations of
$h_{\rm cor}(q)$ from zero indicate the presence of correlations, and
a $q$ dependence of $h_{\rm cor}(q)$ indicates correlation
multifractality [type (ii)].  If only correlation multifractality is
present, $h_{\rm shuf}(q)=0.5$ and $h(q)=0.5+h_{\rm cor}(q)$.  If
both,  distribution multifractality and correlation multifractality
are present, both, $h_{\rm shuf}(q)$ and $h_{\rm cor}(q)$ depend on
$q$.

\subsection{Significance of the results}

In Figs.~\ref{mono1}-\ref{shuf2} we have shown the results of the
MF-DFA for single configurations of long time series.  Now we address
the significance and accuracy of the MF-DFA results for short series. 
How much do the numerically determined exponents $h(q)$ vary from one
configuration (sample series) to the next, and how close are the
average values to the theoretical values?  In other words, how large
are the statistical and systematical deviations of exponents
practically determined by the MF-DFA for finite series?  These
questions are particularly important for short series, where the
statistics is poor.  If the values of $h(q)$ are determined
inaccurately, the multifractal properties will be reported
inaccurately or even false conclusions on multifractal behavior might
be drawn for monofractal series.  

\begin{figure}\centering
\epsfxsize\figsize\epsfbox{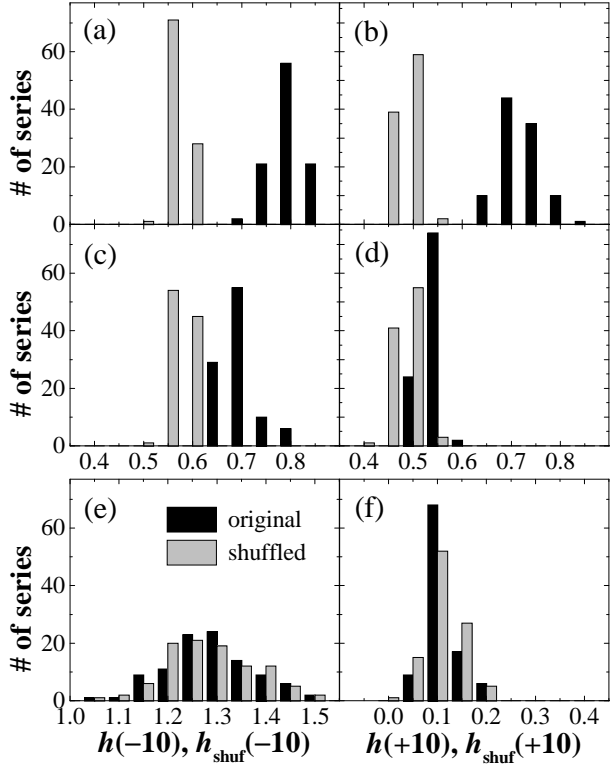}\vspace{1cm}

\parbox{\figsize}{\caption[]{\small
(a) Histograms of the generalized Hurst exponents $h(-10)$ (black bars)
and $h_{\rm shuf}(-10)$ (grey bars) for 100 generated monofractal
series with $H=0.75$.  The exponents have been fitted to MF-DFA2
fluctuation functions in the scaling range $400 < s < 2000$. (b) Same
as (a), but for $h(+10)$ and $h_{\rm shuf}(+10)$.  (c,d) Same as (a,b),
but for the dyadic random cascade model with log-Poisson distribution
and parameters $\lambda = \ln 1$, $\delta = 0.9$, and $\gamma =
-\delta^4/2$.  The corresponding theoretical values are $h(-10)=0.743$
and $h(+10)=0.567$ from Eq.~(\protect{\ref{Hdya}}) for
the original series and $h_{\rm shuf}=0.5$ for the shuffled series. 
From the histogram of $h(+10)$ it would be hard to draw any conclusions
regarding multifractality.  (e,f) Same as (a,b), but for power-law
distributed random numbers with the distribution $P(x)\sim x^{-2}$.
The corresponding theoretical values from Eq.~(\protect{\ref{Hbroa}})
are $h(-10)=1$ and $h(+10)=0.1$ for the original and the shuffled
series.  The length of all series is $L=8192$.  The figure shows that
correlations and multifractality due to correlations (a-d) are
eliminated by the shuffling procedure, while multifractality due to a
broad distribution (e,f) remains.  It further allows to estimate the
statistical fluctuations in the scaling exponents $h(q)$ determined by
the MF-DFA for monofractal (a,b), correlation multifractal (c,d) and
distribution multifractal (e,f) series. \label{his1}}} \end{figure}

To address the significance and accuracy of the MF-DFA results we
generate, for each of the three examples considered already in
Fig.~\ref{shuf1}, 100 series of length $N=2^{13}=8192$ and calculate
$h(-10)$, $h(+10)$, $h_{\rm shuf}(-10)$, and $h_{\rm shuf}(+10)$ for 
each of these series.  The corresponding histograms are shown in
Fig.~\ref{his1}.  For the long-range power-law correlated series with
$H=0.75$ we find the following mean values and standard deviations of
the generalized Hurst exponents:
\begin{eqnarray*} h(-10) = 0.80 \pm 0.03, \quad h_{\rm shuf}(-10) = 
0.56 \pm 0.02, \\ h(+10) = 0.72 \pm 0.04, \quad h_{\rm shuf}(+10) = 
0.48 \pm 0.02. \end{eqnarray*}
The mean values for the original series are rather close to, but not
identical to the theoretical value $H=0.75$.  The mean value for
$q=-10$ is about two standard deviations larger than 0.75, while the
value for $q=+10$ is slightly smaller.  These deviations, though,
certainly {\it cannot} indicate multifractality, since we analyzed
monofractal series.  Instead, they are due to the finite, random
series, where parts of the series have slightly larger and slightly
smaller scaling exponent just by statistical fluctuations.  By
considering negative values of $q$ we focus on the parts with small
fluctuations, which are usually described by a larger scaling exponent
\cite{note3}.  For positive values of $q$ we focus on the  parts with
large fluctuations usually described by a smaller value of $h$.  Thus
for short records we always expect a slight difference between
$h(-10)$ and $h(+10)$ even if the series are monofractal.  If this
difference is weak, one has to be very careful with conclusions about
multifractality.  Practically it is always wise to compare with
generated monofractal series with otherwise similar properties before
drawing conclusions regarding the multifractality of a time series. 
In addition to the statistical fluctuations of the $h$ values, the
average $h(-10)$ is usually determined slightly too large, while
$h(+10)$ is slightly too small.  The same behavior is obtained if the
WTMM method is used instead of the MF-DFA, as we will show in
Subsection V.C.

The same kind of difference is also observed for the average $h_{\rm
shuf}(-10)$ and $h_{\rm shuf}(+10)$ values.  After all correlations
have been destroyed by the shuffling, $h_{\rm shuf}=0.5$ is expected
since the probability density is Gaussian with all finite moments. 
The deviations from $h_{\rm shuf}=0.5$ we observe for the finite
random series are characteristic for monofractal series of this length
($N=8192$).  Only for the second moment we obtain $h_{\rm shuf}(2) =
0.5$ exactly if a sufficient number of series is considered. 

For multifractal series generated from the dyadic random cascade
model, Fig.~\ref{his1}(c,d) shows the histograms of the scaling
exponents $h(-10)$, $h(+10)$, $h_{\rm shuf}(-10)$, and $h_{\rm
shuf}(+10)$.  Their averages and standard deviations,
\begin{eqnarray*} h(-10) = 0.69 \pm 0.04, \quad h_{\rm shuf}(-10) = 
0.57 \pm 0.02, \\ h(+10) = 0.54 \pm 0.02, \quad h_{\rm shuf}(+10) = 
0.48 \pm 0.02. \end{eqnarray*}
have to be compared with the theoretical values from Eq.~(\ref{Hdya}),
$h(-10)=0.743$ and $h(+10)=0.567$. 
Surprisingly, the mean $h(-10)$ is smaller than the theoretical value
in this example, but for the mean $h(+10)$ the deviation is similar to
the deviation observed for the monofractal data in the previous
example.  Again, similar results are obtained with the WTMM method. 
For the shuffled series, the mean generalized Hurst exponents are
practically identical to those for the shuffled monofractal series
[the average $h_{\rm shuf}(-10)$ is larger by half the standard
deviation], and both are evidently consistent with monofractal
uncorrelated behavior, $h(q)=0.5$, as discussed above.  Hence, the
series from the dyadic random cascade model show no signs of
distribution multifractality and are characterized by correlation
multifractality only.

The histograms of the scaling exponents for our last example, the
power-law distributed random numbers with $P(x) \sim x^{-2}$, are
shown in Fig.~\ref{his1}(e,f).  The corresponding mean values and
standard deviations,
\begin{eqnarray*} h(-10) = 1.24 \pm 0.09, \quad h_{\rm shuf}(-10) = 
1.26 \pm 0.09, \\ h(+10) = 0.11 \pm 0.03, \quad h_{\rm shuf}(+10) = 
0.11 \pm 0.04. \end{eqnarray*}
show obviously no differences between original and shuffled series as
expected for uncorrelated series.  This indicates that the
multifractality is due to the broad probability density function
only.  The values have to be compared with $h(-10) = 1$ and $h(+10) =
0.1$ from Eq.~(\ref{Hbroa}).  As usual, the average value of $h(-10)$
is too large because we analyzed short series.

\section{Comparison to the wavelet transform modulus maxima method}

\subsection{Brief description of the wavelet transform modulus maxima
method}

The wavelet transform modulus maxima (WTMM) method \cite{wtmm} is a
well-known method to investigate the multifractal scaling properties
of fractal and self-affine objects in the presence of
nonstationarities.  It is an application of the wavelet transform with
continuous basis functions.  One defines the wavelet-transform of a
series $x_k$ of length $N$ by
\begin{equation} W(n,s) = {1 \over s}
\sum_{k=1}^N x_k \, \psi[(k-n)/s]. \label{eq10}\end{equation}
Note that in this case the series $x_k$ are analyzed directly instead
of the profile $Y(i)$ defined in Eq.~(\ref{profile}).  Here, the
function $\psi(x)$ is the analyzing wavelet and $s$ is, as above, the
scale parameter.  The wavelet is chosen orthogonal to the possible
trend.  If the trend can be represented by a polynomial, a good choice
for $\psi(x)$ is the $m$-th derivative of a Gaussian, $\psi^{(m)}(x) =
d^m(e^{-x^2/2})/dx^m$.  This way, the transform eliminates trends up
to $(m-1)$th order.

Now, instead of averaging over all values of $W(n,s)$, one averages,
within the modulo-maxima method, only the local maxima of $\vert
W(n,s) \vert$.  First, one determines for a given scale $s$, the
positions $n_i$ of the local maxima of $\vert W(n,s) \vert$ as
function of $n$, so that $\vert W(n_i-1,s) \vert < \vert W(n_i,s)
\vert \ge \vert W(n_i+1,s) \vert$ for $i = 1,\ldots,i_{\rm max}$. 
Then one sums up the $q$th power of these maxima,
\begin{equation} Z(q,s) = \sum_{i=1}^{i_{\rm max}} \vert W(n_i,s)
\vert^q. \label{eq11a}\end{equation}
The reason for this maxima procedure is that the absolute wavelet
coefficients $\vert W(n,s) \vert$ can become arbitrarily small.  The
analyzing wavelet $\psi(x)$ must always have positive values for some
$x$ and negative values for other $x$, since it has to be orthogonal
to possible constant trends.  Hence there are always positive and
negative terms in the sum (\ref{eq10}), and these terms might cancel.
If that happens, $\vert W(n,s) \vert$ can become close to zero.
Since such small terms would spoil the calculation of negative moments
in Eq.~(\ref{eq11a}), they have to be eliminated by the maxima
procedure.  In the MF-DFA, the calculation of the variances
$F^2(s,\nu)$ in Eq.~(\ref{fsdef}), i.~e. the deviations from the fits,
involves only positive terms under the summation.  The variances
cannot become arbitrarily small, and hence no maximum procedure is
required for series with compact support. 

In addition, the MF-DFA
variances will always increase if the segment length $s$ is increased,
because the fit will always be worse for a longer segment.  In the
WTMM method, in contrast, the absolute wavelet coefficients $\vert
W(n,s) \vert$ need not increase with increasing scale $s$, even if
only the local maxima are considered. The values $\vert W(n,s) \vert$
might become smaller for increasing $s$ since just more (positive and
negative) terms are included in the summation (\ref{eq10}), and these
might cancel even better.  Thus, an additional supremum
procedure has been introduced in the WTMM method in order to keep the
dependence of $Z(q,s)$ on $s$ monotonous: If, for a given scale $s$, a
maximum at a certain position $n_i$ happens to be smaller than a
maximum at $n'_i \approx n_i$ for a lower scale $s' < s$, then
$W(n_i,s)$ is replaced by $W(n'_i,s')$ in Eq.~(\ref{eq11a}).  There
is no need for such a supremum procedure in the MF-DFA.

Often, scaling behavior is observed for $Z(q,s)$, and scaling
exponents $\hat{\tau}(q)$ can be defined that describe how $Z(q,s)$
scales with $s$,
\begin{equation} Z(q,s) \sim s^{\hat{\tau}(q)}.
\label{eq11b}\end{equation}
The exponents $\hat{\tau}(q)$ characterize the multifractal properties
of the series under investigation, and theoretically they are
identical to the $\tau(q)$ defined in Eq.~(\ref{Zq}) \cite{wtmm} and
related to $h(q)$ in Eq.~(\ref{tauH}).

\subsection{Examples for series with nonstationarities}

\begin{figure}\centering
\epsfxsize\figsize\epsfbox{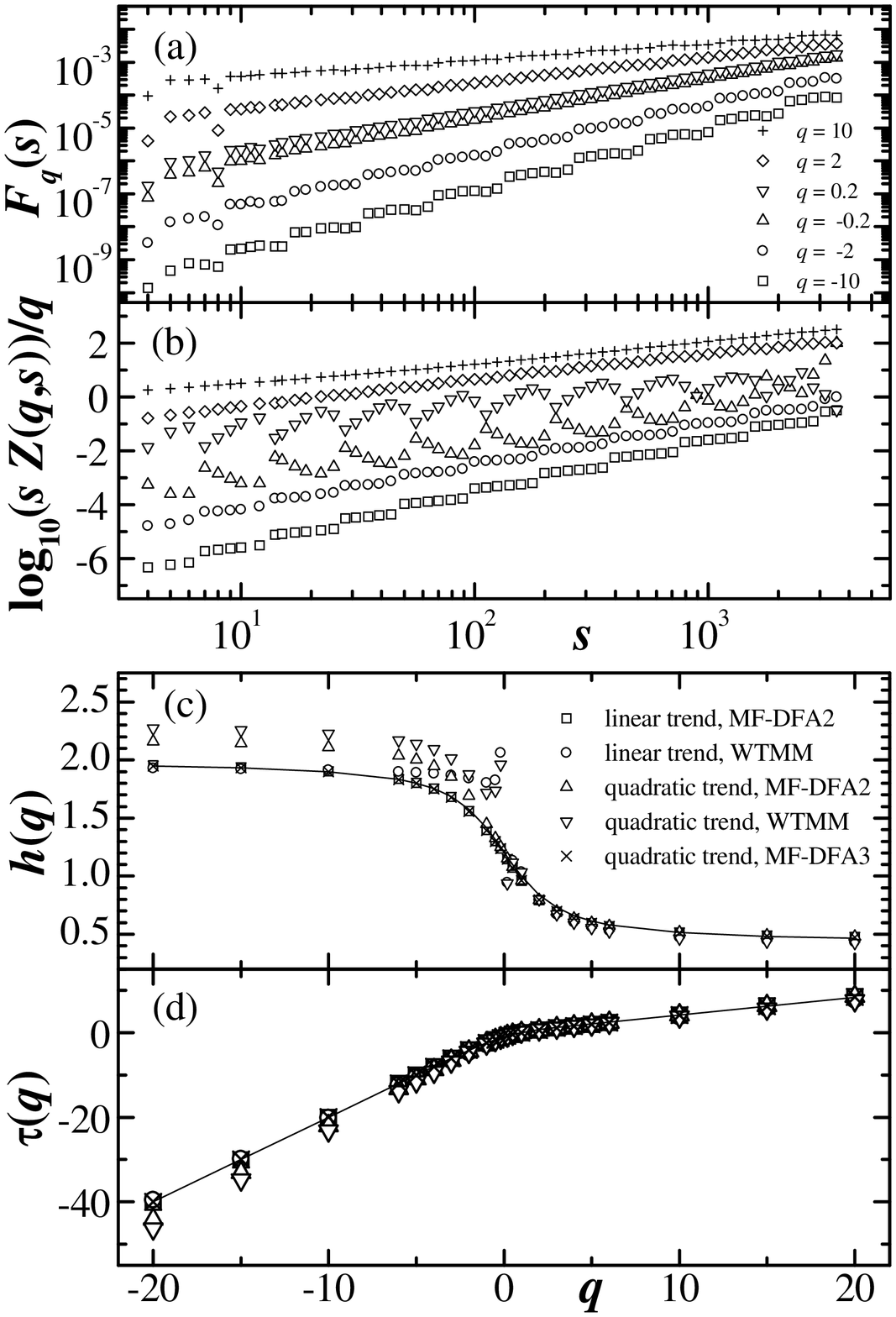}\vspace{1cm}

\parbox{\figsize}{\caption[]{\small
(a) The MF-DFA2 fluctuation functions $F_q(s)$ are shown versus the
scale $s$ in log-log plots for the binomial multifractal model with
$a=0.75$ and an additional linear trend $x_k \to x_k + k/500 L$. (b)
The scaled WTMM partition functions $[s Z(q,s)]^{1/q}$ are shown for
the same series and the same values of  $q$.  The symbols are the same
as for Fig.~1.  (c) The $q$ dependence of the generalized Hurst
exponent $h(q)$ for the generated series with  linear trend for the
MF-DFA2 ($\Box$) and the second order WTMM ({\Large$\circ$}) methods.
Corresponding results for a binomial multifractal with an additional
quadratic trend are also included for MF-DFA2 ($\bigtriangleup$) and
second order WTMM ($\bigtriangledown$) methods.  The quadratic trend
causes deviations from  the line indicating the theoretical values
[obtained from Eq.~(\protect\ref{Hbin})], which disappear if MF-DFA3
is employed ($\times$).   The values of $h(q)$ have been determined by
fits to $F_q(s)$ and $Z(q,s)$  in the regime $50 < s < 2000$.  The
relation $h(q) = [\tilde{\tau}(q) + 1]/q$ from Eq.~(\ref{tauH}) has
been used to convert the exponent $\tilde{\tau}(q)$ from
Eq.~(\ref{eq11b}) into $h(q)$.  (d) The $q$ dependence of $\tau(q)$.
\label{bin2}}}\end{figure}

Since the WTMM method has been developed to analyze multifractal
series with nonstationarities, such as trends or spikes, we will
compare its performance with the performance of the MF-DFA for such
nonstationary series.  In Fig.~\ref{bin2} the MF-DFA fluctuation
function $F_q(s)$ and its scaling behavior are compared with the
rescaled WTMM partition sum $Z(q,s)$ for the binomial multifractal
described in Subsection III.B.  To test the detrending capability of
both methods, we have added linear as well as quadratic trends to the
generated multifractal series.  The trends are removed by both
methods, if a sufficiently high order of detrending is employed.  The
deviations from the theoretical values of the scaling exponents $h(q)$
[given by Eq.~(\ref{Hbin})] are of similar size for the MF-DFA and the
WTMM method.  Thus, the detrending capability and the accuracy of both
methods is equivalent.

We also obtain similar results for a monofractal long-range correlated
series with additional spikes (outliers) that consist of large random
numbers and replace a small fraction of the original series in
randomly chosen positions.  The spikes lead to multifractality on small
scales $s$, while the series remains monofractal on large scales. 
Thus, the effects of the spikes are eliminated neither by the WTMM
method nor by the MF-DFA, but both methods again give rather equivalent
results.

\subsection{Significance of the results}

The last problem we address is a comparison of the significance of
the results obtained by the MF-DFA and the WTMM method. The
significance of the MF-DFA results has already been discussed in
detail in Subsection IV.B.  Here we will compare the significance
of both methods for short and long series.

\begin{figure}\centering
\epsfxsize\figsize\epsfbox{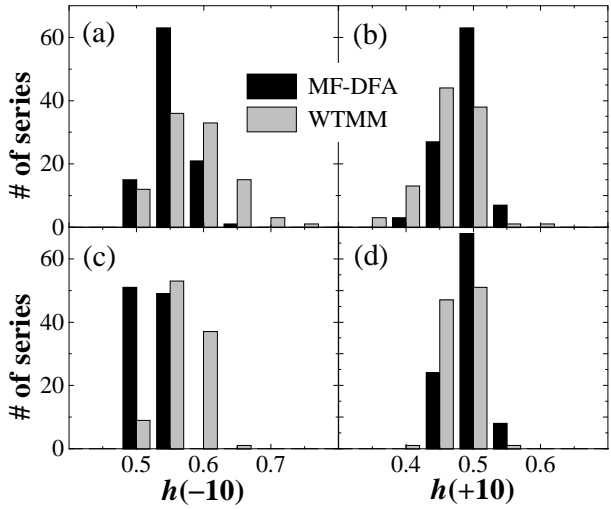}\vspace{1cm}

\parbox{\figsize}{\caption[]{\small
(a) Histograms of the generalized Hurst exponents $h(-10)$ for 100
random uncorrelated series with $H=0.5$.  The exponents have been
fitted to MF-DFA2 fluctuation functions $F_{-10}(s)$ (black bars) in
the scaling range $40 < s < 2000$ and to WTMM results $Z(-10,s)$ (grey
bars) in the scaling range $5 < s < 250$.  The length of the series is
$L=8192$.  The relation $h(q) = [\hat{\tau}(q) + 1]/q$ from
Eq.~(\ref{tauH}) has been used to convert the exponent
$\hat{\tau}(q)$ from Eq.~(\ref{eq11b}) into $h(q)$.  (b) Same as
(a), but for $h(+10)$.  (c,d) Same as (a,b), but for longer series
($L=65536$), where statistical fluctuations are reduced.  The figure
shows that the MF-DFA seems to give slightly more reliable results
than the WTMM method for short series and negative moments ($q=-10$),
see (a).  In the other cases, the performance of both methods is
similar. \label{his2}}}\end{figure}

We begin with the significance of the results for random series
involving neither correlations nor a broad distribution [as in
Fig.~\ref{mono1}(b)].  Fig.~\ref{his2} shows the distribution of the
multifractal Hurst exponents $h(-10)$ and $h(+10)$ calculated by the
MF-DFA as well as by the WTMM using the relation $h(q) = [\hat{\tau}
(q) + 1]/q$ based on Eq.~(\ref{tauH}).  Similar to the results
presented in Fig.~\ref{his1}, we have analyzed 100 generated series
of uncorrelated random numbers.  In addition, we compare the results
for the (relatively short) series length $N=2^{13}=8192$ and for
$N=2^{16}=65532$.  Ideally, both, $h(-10)$ and $h(+10)$, should be
equal to the Hurst exponent of the uncorrelated monofractal series,
$H=0.5$.  The histograms show that similar deviations as well as
remarkable fluctuations of the exponents occur for both methods, as
discussed in Subsection IV.B for the MF-DFA.   We find the following
mean values and standard deviations,
\begin{eqnarray*} h(-10) =  \cases{ 0.55 \pm 0.03 & for MF-DFA
($N=8k$) \cr 0.52 \pm 0.02 & for MF-DFA ($N=64k$)  \cr 0.58 \pm 0.05
& for WTMM ($N=8k$) \cr 0.56 \pm 0.03 & for WTMM ($N=64k$)} \\ {\rm
and } \quad h(+10) = \cases{ 0.49 \pm 0.03 & for MF-DFA ($N=8k$) \cr
0.49 \pm 0.02 & for MF-DFA ($N=64k$)  \cr 0.46 \pm 0.04 & for WTMM
($N=8k$) \cr 0.48 \pm 0.02 & for WTMM ($N=64k$)}. \end{eqnarray*}
As already discussed in Subsection IV.B, the deviations of the average
$h(q)$ values from $H=0.5$ do not indicate multifractality.  For the
WTMM method and short series, one has to be very careful in order not
to draw false conclusions from results like $h(-10)=0.58$ and
$h(+10)=0.46$.  The corresponding results of the MF-DFA are closer to
the theoretical value.

\begin{figure}\centering
\epsfxsize\figsize\epsfbox{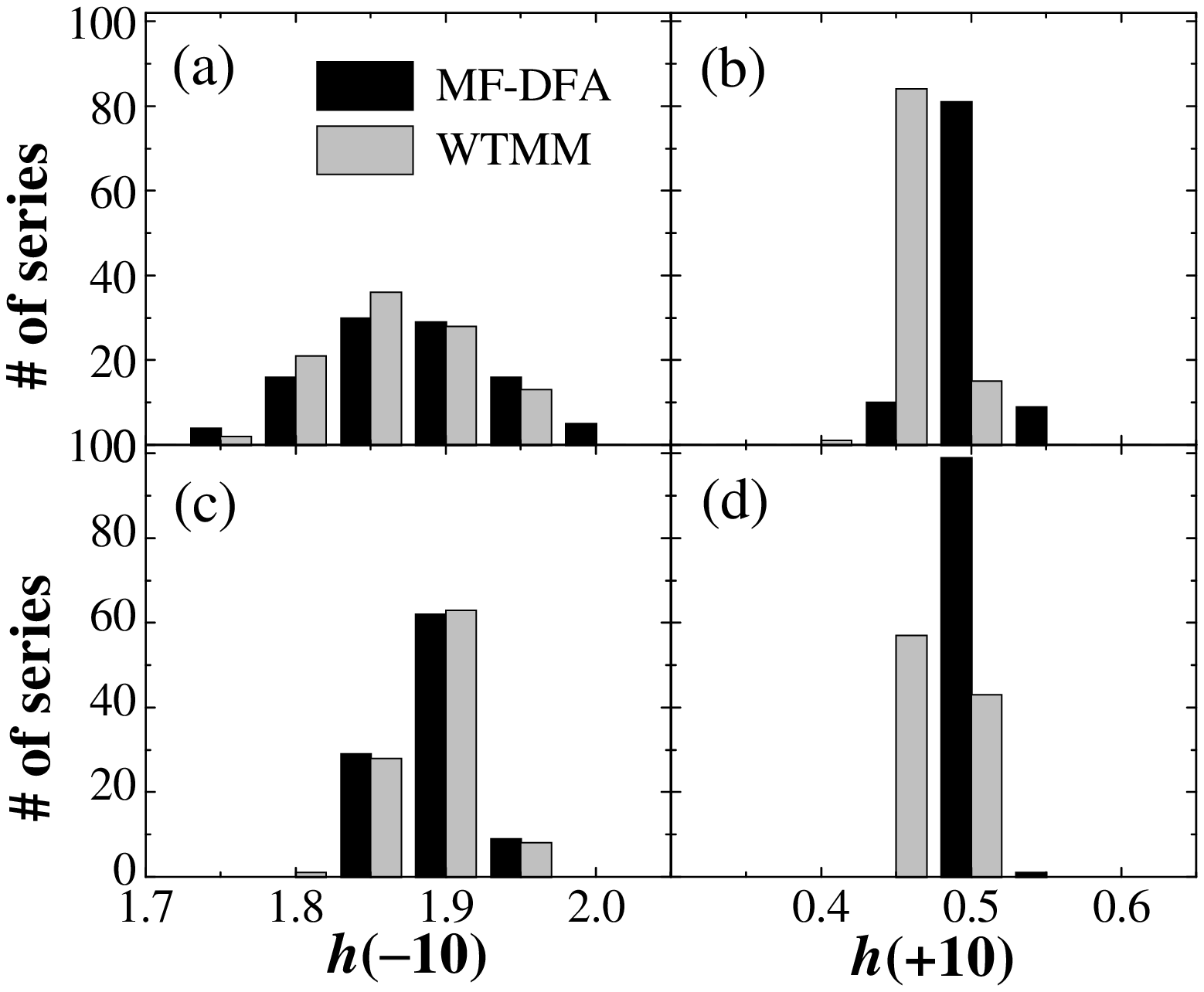}\vspace{1cm}

\parbox{\figsize}{\caption[]{\small
Same as Fig.~\protect\ref{his2} for the binomial model with $a=0.75$.
The theoretical values of the generalized Hurst exponents are $h(-10)
= 1.90$ and $h(+10) = 0.515$ according to Eq.~(\protect\ref{Hbin}). 
The figure shows that our findings regarding the performance of the
MF-DFA and WTMM methods for uncorrelated monofractal series in 
Fig.~\protect\ref{his2} also hold for multifractal series.
\label{his3}}}\end{figure}

Figure \ref{his3} shows the distribution of the multifractal scaling
exponents $h(-10)$ and $h(+10)$ calculated for generated multifractal
series from the binomial model with $a=0.75$ described in Subsection
III.B.  Like for Fig.~\ref{his2}, 100 generated series have been
analyzed for each of the histograms.  Now, the differences between the
distributions of $h(-10)$ and $h(+10)$ are much larger, indicating
multifractality.  We find
\begin{eqnarray*} h(-10) =  \cases{ 1.88 \pm 0.06 & for MF-DFA
($N=8k$) \cr 1.89 \pm 0.03 & for MF-DFA ($N=64k$)  \cr 1.86 \pm 0.05
& for WTMM ($N=8k$) \cr 1.89 \pm 0.02 & for WTMM ($N=64k$)} \\ {\rm
and } \quad h(+10) = \cases{ 0.50 \pm 0.02 & for MF-DFA ($N=8k$) \cr
0.51 \pm 0.01 & for MF-DFA ($N=64k$)  \cr 0.46 \pm 0.01 & for WTMM
($N=8k$) \cr 0.47 \pm 0.01 & for WTMM ($N=64k$)}. \end{eqnarray*}
These values must be compared with the theoretical values $h(-10) =
1.90$ and $h(+10) = 0.515$ from Eq.~(\ref{Hbin}).  Again, the MF-DFA
results turn out to be slightly more significant than the WTMM
results.  The MF-DFA seems to have slight advantages for negative $q$
values and short series, but in the other cases the results of the
two methods are rather equivalent.  Besides that, the main advantage
of the MF-DFA method compared with the WTMM method lies in the
simplicity of the MF-DFA method.

\section{Conclusion}

We have generalized the DFA, widely recognized as a method to analyze
the (mono-) fractal scaling properties of nonstationary time series.
The MF-DFA method allows a reliable multifractal characterization of
multifractal nonstationary time series.  The implementation of the
new method is not more difficult than that of the conventional DFA,
since just one additional step, a $q$ dependent averaging procedure,
is required.  We have shown for stationary signals that the
generalized (multifractal) scaling exponent $h(q)$ for series with
compact support is directly related to the exponent $\tau(q)$ of the
standard partition function-based multifractal formalism.  Further, we
have shown in several examples that the MF-DFA method can reliably
determine the multifractal scaling behavior of the time series,
similar to the WTMM method which is a more complicated procedure for
this purpose.  For short series and negative moments, the significance
of the results for the MF-DFA seems to be slightly better than for the
WTMM method.

Contrary to the WTMM method, the MF-DFA method as described in Subsection
II.A requires series of compact support,  because the averaging
procedure in Eq.~(\ref{fdef}) will only work if $F^2(s,\nu) > 0$ for
all segments $\nu$.  Although most time series will fulfill this
prerequisite, it can be overcome by a modification of the MF-DFA
technique in order to analyze data with fractal support:  We restrict
the sum in Eq.~(\ref{fdef}) to the local maxima, i.~e. to those terms
$F^2(s,\nu)$ that are larger than the terms $F^2(s,\nu-1)$ and
$F^2(s,\nu+1)$ for the neighboring segments.  By this restriction all
terms $F^2(s,\nu)$ that are zero or very close to zero will be
disregarded, and series with fractal support can be analyzed.  The
procedure reminds slightly of the modulus maxima procedure in the WTMM
method (see Subsection V.A).  There is no need, though, to employ a
continuously sliding window or to calculate the supremum over all
lower scales for the MF-DFA, since the variances $F^2(s,\nu)$, which
are determined by the deviations from the fit, will always increase
when the segment size $s$ is increased.  In the maxima MF-DFA
procedure the generalized Hurst exponent $h(q)$ defined in
Eq.~(\ref{Hq}) will depend on $q$ and even diverge for $q \to 0$ for
monofractal series with non-compact support.  Thus, it is more
appropriate to consider the scaling exponent $\tau(q)$, calculating
\begin{equation} \sum_{F^2(s,\nu-1) < F^2(s,\nu) \ge F^2(s,\nu+1)}
[F^2(s,\nu)]^{q/2} \sim s^{\tau(q)}. \end{equation}
This extended MF-DFA procedure will also be applicable for data with
fractal support.

In a later work we will apply the MF-DFA method to a range of
physiological and meteorological data.

\subsection*{Acknowledgements}

We would like to thank Yosef Ashkenazy for useful discussions and the
Deutsche Akademischer Austauschdienst, the Deutsche
Forschungsgemeinschaft, the German Israeli Foundation, and the Minerva
Foundation for financial support.

\begin{thebibliography} {50}

\bibitem{Peng94} C.-K. Peng, S. V. Buldyrev, S. Havlin, M. Simons, H.
 E. Stanley, and A. L. Goldberger, Phys. Rev. E {\bf 49}, 1685
 (1994);
S. M. Ossadnik, S. B. Buldyrev, A. L. Goldberger, S. Havlin, R.N.
 Mantegna, C.-K. Peng, M. Simons, and H.E. Stanley, Biophys. J. {\bf
 67}, 64 (1994).
\bibitem{taqqu} M. S. Taqqu, V. Teverovsky, and W. Willinger,
 Fractals {\bf 3}, 785 (1995).
\bibitem{physa} J. W. Kantelhardt, E. Koscielny-Bunde, H. H. A. Rego,
 S. Havlin, and A. Bunde, Physica A {\bf 295}, 441 (2001).
\bibitem{kunhu} K. Hu, P. Ch. Ivanov, Z. Chen, P. Carpena, and H. E.
 Stanley, Phys. Rev. E {\bf 64}, 011114 (2001).
\bibitem{kunhu1} Z. Chen, P. Ch. Ivanov, K. Hu, and H. E. Stanley,
 Phys. Rev. E {\bf 65}, xxxx (April 2002), preprint physics/0111103.

\bibitem{dns} S. V. Buldyrev, A. L. Goldberger, S. Havlin, R. N.
 Mantegna, M. E. Matsa, C.-K. Peng, M. Simons, and H. E. Stanley,
 Phys. Rev. E {\bf 51}, 5084 (1995);
S. V. Buldyrev, N. V. Dokholyan, A. L. Goldberger,
 S. Havlin, C.-K. Peng, H. E. Stanley, and G. M. Viswanathan,
 Physica A {\bf 249}, 430 (1998).
\bibitem{herz} C.-K. Peng, S. Havlin, H. E. Stanley,
 and A. L. Goldberger, Chaos {\bf 5}, 82 (1995);
P. Ch. Ivanov, A. Bunde, L. A. N. Amaral, S. Havlin, J.
 Fritsch-Yelle, R. M. Baevsky, H. E. Stanley, and A. L. Goldberger,
 Europhys. Lett. {\bf 48}, 594 (1999);
Y. Ashkenazy, M. Lewkowicz, J. Levitan, S. Havlin,
 K. Saermark, H. Moelgaard, P. E. B. Thomsen, M. Moller, U. Hintze,
 and H. V. Huikuri, Europhys. Lett. {\bf 53}, 709 (2001);
Y. Ashkenazy, P. Ch. Ivanov, S. Havlin, C.-K. Peng, A. L. Goldberger,
 and H. E. Stanley, Phys. Rev. Lett. {\bf 86}, 1900 (2001).
\bibitem{PRL00} A. Bunde, S. Havlin, J. W. Kantelhardt, T. Penzel,
 J.-H. Peter, and K. Voigt, Phys. Rev. Lett. {\bf 85}, 3736 (2000).
\bibitem{neuron} S. Blesic, S. Milosevic, D. Stratimirovic, and
 M. Ljubisavljevic, Physica A {\bf 268}, 275 (1999);
S. Bahar, J. W. Kantelhardt, A. Neiman, H. H. A. Rego, D. F. Russell,
 L. Wilkens, A. Bunde, and F. Moss, Europhys. Lett. {\bf 56}, 454
 (2001).
\bibitem{gait} J. M. Hausdorff, S. L. Mitchell, R. Firtion, C.-K.
 Peng, M. E. Cudkowicz, J. Y. Wei, and A. L. Goldberger, J. Appl.
 Physiology {\bf 82}, 262 (1997).
\bibitem{wetter} E. Koscielny-Bunde, A. Bunde, S. Havlin, H.E. Roman,
 Y. Goldreich, and H.-J. Schellnhuber, Phys. Rev. Lett. {\bf 81}, 729
 (1998).
K. Ivanova and M. Ausloos, Physica A {\bf 274}, 349 (1999);
P. Talkner and R.O. Weber, Phys. Rev. E {\bf 62}, 150 (2000).
\bibitem{cloud} K. Ivanova, M. Ausloos, E. E. Clothiaux, and T. P.
 Ackerman, Europhys. Lett. {\bf 52}, 40 (2000).
\bibitem{malamudjstatlaninfer1999} B. D. Malamud and D. L. Turcotte,
 J. Stat. Plan. Infer. {\bf 80}, 173 (1999).
\bibitem{Alados2000} C. L. Alados and M. A. Huffman,
 Ethnology {\bf 106}, 105 (2000).
\bibitem{economics} R. N. Mantegna and H. E. Stanley, {\it An
 Introduction to Econophysics} (Cambridge University Press,
 Cambridge, 2000);
Y. Liu, P. Gopikrishnan, P. Cizeau, M. Meyer, C.-K. Peng, and H. E.
 Stanley, Phys. Rev. E {\bf 60}, 1390 (1999);
N. Vandewalle, M. Ausloos, and P. Boveroux, Physica A {\bf 269}, 170
 (1999).
\bibitem{fest} J. W. Kantelhardt, R. Berkovits, S. Havlin, and A.
 Bunde, Physica A {\bf 266}, 461 (1999);
N. Vandewalle, M. Ausloos, M. Houssa, P. W. Mertens, and M. M. Heyns,
 Appl. Phys. Lett. {\bf 74}, 1579 (1999).

\bibitem{feder88} J. Feder, {\em Fractals} (Plenum Press, New York,
 1988).
\bibitem{barabasi} A.-L. Barab\'asi and T. Vicsek, Phys. Rev. A
 {\bf 44}, 2730 (1991).
\bibitem{peitgen}  H.-O. Peitgen, H. J\"urgens, and D. Saupe, {\it
 Chaos and Fractals} (Springer-Verlag, New York, 1992), Appendix B.
\bibitem{bacry01} E. Bacry, J. Delour, and J. F. Muzy,
 Phys. Rev. E {\bf 64}, 026103 (2001).
\bibitem{wtmm} J. F. Muzy, E. Bacry, and A. Arneodo,
 Phys. Rev. Lett. {\bf 67}, 3515 (1991);
J. F. Muzy, E. Bacry, and A. Arneodo,
 Int. J. Bifurcat. Chaos {\bf 4}, 245 (1994);
A. Arneodo, E. Bacry, P. V. Graves, and J. F. Muzy,
 Phys. Rev. Lett. {\bf 74}, 3293 (1995);
A. Arneodo et al. in: {\it The science of disaster: climate
 disruptions, market crashes, and heart attacks}, ed. by A. Bunde and
 H. J. Schellnhuber (Springer-Verlag, Berlin, 2002).

\bibitem{note2}  The value of $h(0)$, which corresponds to the limit
$h(q)$ for $q \to 0$ for time series with compact support, cannot be
determined directly using the averaging procedure in Eq.~(\ref{fdef})
because of the diverging exponent.  Instead, a logarithmic averaging
procedure has to be employed,
%
\begin{equation} F_0(s) \equiv \exp \left\{ {1 \over 4 N_s}
\sum_{\nu=1}^{2 N_s} \ln \left[F^2(s,\nu)\right] \right\}
\sim s^{h(0)}. \end{equation}
%
Note that $h(0)$ cannot be defined for time series with fractal
support, where $h(q)$ diverges for $q \to 0$.

\bibitem{note3}  For the maximum scale $s=N$ the fluctuation function
$F_q(s)$ is independent of $q$, since the sum in Eq.~(\ref{fdef}) runs
over only two identical segments ($N_s \equiv [N/s] = 1$).  For
smaller scales $s \ll N$ the averaging procedure runs over several
segments, and the average value $F_q(s)$ will be dominated by the
$F^2(s,\nu)$ from the segments with small (large) fluctuations if $q
< 0$ ($q > 0$).  Thus, for $s \ll N$, $F_q(s)$ with $q < 0$ will
be smaller than $F_q(s)$ with $q > 0$, while both become equal for
$s=N$.  Hence, if we assume an homogeneous scaling behavior of $F_q(s)$
following Eq.~(\ref{Hq}), the slope $h(q)$ in a log-log plot of
$F_q(s)$ with $q < 0$ versus $s$ must be larger than the
corresponding slope for $F_q(s)$ with $q > 0$.  Thus, $h(q)$ for $q
< 0$ will usually be larger than $h(q)$ for $q > 0$.

\bibitem{makse96} H. A. Makse, S. Havlin, M. Schwartz, H. E. Stanley,
 Phys. Rev. E {\bf 53}, 5445 (1996).
\bibitem{arneodo98} A. Arneodo, E. Bacry, and J. F. Muzy, J. Math.
 Phys. {\bf 39}, 4142 (1998); A. Arneodo, S. Manneville, and J. F. Muzy,
 Europhys. J. B {\bf 1}, 129 (1998).
\bibitem{PREprep} Y. Ashkenazy, S. Havlin. P. Ch. Ivanov, C. K. Peng,
 V. Schulte-Frohlinde, and H. E. Stanley, cond-mat/0111396 (unpublished).

\bibitem{shle87} M. F. Shlesinger, B. J. West, and J. Klafter,
 Phys. Rev. Lett. {\bf 58}, 1100 (1987).
\bibitem{havlin99} S. Havlin and Y. Ben Avraham, {\it Diffusion and
 Reactions in Fractals and Disordered Systems}, (Cambridge University
 Press, Cambridge, 2000), p. 48, and references therein.
\bibitem{newprep} N. Scafetta and P. Grigolini, cond-mat/0202008
 (unpublished).
\bibitem{jaffard} S. Jaffard,  Probab. Theory Rel. {\bf 114}, 207 
 (1999).
\bibitem{nakao} H. Nakao, Phys. Lett. A {\bf 266}, 282 (2000).

\bibitem{note4}  The example of the binomial multifractal series
(Subsection III.B) can also show multifractality due to a broad
probability density function for the values $x_k$.  If the parameter
$a$ is chosen to be very close to one or if very long series are
considered, corresponding to large values of $n_{\rm max}$, the
minimum value in the series, $(1-a)^{n_{\rm max}}$, will be very small
compared with the maximum value $a^{n_{\rm max}}$ [see
Eq.~(\ref{bindef})].  In this case the log-binomial probability
density function will become broad, approaching a log-normal form.
Since the scaling behavior of uncorrelated log-normal distributed
series corresponds to the multifractal scaling behavior observed in
the example of uncorrelated power-law distributed series with
$\alpha=2$ (see Subsection III.D), distribution multifractality [type
(i)] will occur in addition to the correlation multifractality [type
(ii)].  For the series with $a = 0.75$ and $N=8192$ ($n_{\rm max} =
13$) considered in Subsection III.B, we observe only type (ii)
multifractality caused by long-range correlations.

\bibitem{note5}  The ratio $F_q(s)/F_q^{\rm shuf}(s)$ can also be used
to eliminate systematic deviations from the expected power-law scaling
behavior that occur at very small scales $s<10$ especially for small
values of $q$ (see Figs.~1-3).  A similar procedure has already been
introduced for the conventional DFA recently (see Subsection 3.1 of
\cite{physa}).  Since the deviations are systematic for the MF-DFA
method, they occur in both, $F_q(s)$ and $F_q^{\rm shuf}(s)$, and they
should cancel in the ratio.

\end{thebibliography}
\end{document}